\documentclass[12pt]{article}
\usepackage{hyperref}
\usepackage{amsmath}
\usepackage{graphicx}
\usepackage{amssymb} 
 \usepackage{xcolor}
\definecolor{myblue}{rgb}{0.14,0.11,0.49}
\definecolor{myred}{rgb}{0.74,0.22,0.15}
\definecolor{mygreen}{rgb}{0.05,0.52,0.42}
\definecolor{myyellow}{rgb}{0.96,0.92,0.13}
\definecolor{myorange}{rgb}{1,0.61,0.36}
\definecolor{mypurple}{rgb}{0.71,0.02,1}
\newcommand{\Couleur}[1]{\textcolor{myblue}{#1}} 
\def\be{\begin{equation}}
\def\ee{\end{equation}}
\def\bea{\begin{eqnarray}}
\def\eea{\end{eqnarray}}
\def\bi{\begin{itemize}}
\def\ei{\end{itemize}}
\def\noi{\noindent}

\date{}

\title{On reference frames and the definition of space\\ in a general spacetime}

\author{Mayeul Arminjon\\
\small\it Laboratory ``Soils, Solids, Structures, Risks'', 3SR\\ \small\it (CNRS and Universit\'es de Grenoble: UJF, Grenoble-INP),\\\small\it BP 53, F-38041 Grenoble cedex 9, France.}

\begin{document}
\maketitle

\begin{abstract} 
\noi 
First, we review local concepts defined previously. A (local) reference frame \Couleur{$\mathrm{F}$} can be defined as an equivalence class of admissible spacetime charts (coordinate systems) having a common domain \Couleur{$\mathrm{U}$} and exchanging by a spatial coordinate change. The associated (local) physical space is made of the world lines having constant space coordinates in any chart of the class. Second, we introduce new, global concepts. The data of a non-vanishing global vector field \Couleur{$\,v\,$} defines a global ``reference fluid". The associated global physical space is made of the maximal integral curves of that vector field. Assume that, in any of the charts which make some reference frame \Couleur{$\mathrm{F}$}: (i) any of those integral curves \Couleur{$l$} has constant space coordinates \Couleur{$x^j$}, and (ii) the mapping \Couleur{$l\mapsto (x^j)$} is one-to-one. In that case, the local space can be identified with a part (an open subset) of the global space.
\end{abstract} 

\section{Introduction}
\vspace{2mm}

A reference frame, in a broad physical sense, is a three-dimensional network of observers equipped with clocks and meters. To any reference frame one should be able to associate some three-dimensional {\it space,} in which the observers of the network are by definition at rest (even though their mutual distances may depend on time). Clearly, both notions are fundamental ones for physics. In Newtonian physics, the consideration is in general (though not always \cite{Moreau1945, A8}) restricted to reference frames that are {\it rigid} with respect to the invariant Euclidean space metric. The same restriction is used in special relativity: there, one considers mainly the {\it inertial frames,} each of which is rigid with respect to the {\it spatial metric} in the considered reference frame.\\

In the relativistic theories of gravitation, the main object is the spacetime metric, which is a field, i.e. it depends on the spacetime position. Hence, rigid reference frames are not relevant any more. The relevant notion is that of a {\it reference fluid.} A three-dimensional network of observers is defined by {\it a time-like vector field \Couleur{$v$} on spacetime} \cite{Cattaneo1958, Massa1974a, Massa1974b, Mitskievich1996}: \Couleur{$v$} is the unit tangent vector field to the world lines of the observers belonging to the network. However, in the general-relativistic literature, it is often implicit that a reference frame can be defined from the data of a {\it coordinate system} (or {\it chart}); see e.g. Landau \& Lifshitz \cite{L&L} and M\o ller \cite{Moeller1952}. The link with the definition by a 4-velocity vector field \Couleur{$v$} was done by Cattaneo \cite{Cattaneo1958}. Namely, any admissible chart on the spacetime, \Couleur{$\chi: X \mapsto (x^\mu) \ (\mu =0,...,3)$}, defines a unique reference fluid, given by its four-velocity field \Couleur{$v$}: the components of \Couleur{$v$} in the chart \Couleur{$\chi$} are 
\be\label{Vmu}
\Couleur{v^0\equiv \frac{1}{\sqrt{g_{00}}}, \qquad v^j=0 \quad (j=1,2,3)}.
\ee
The vector (\ref{Vmu}) is invariant under the ``internal changes''
\be\label{internal-change}
\Couleur{x'^0=\phi ((x^\mu)),\quad x'^k=\phi ^k((x^j)) \quad (j,k=1,2,3)}.
\ee
We note, however, that this is valid only within the domain of definition of the chart \Couleur{$\chi$} --- an open subset \Couleur{$\mathrm{U}$} of the whole spacetime manifold \Couleur{$\mathrm{V}$}.\\

The notion of the {\it space associated} with a reference fluid/network was missing in the general-relativistic literature. However, it is apparent in experimental or observational papers that one cannot dispense with the notion of a spatial position (of any part of the experimental apparatus and the observed system). In the absence of a definite concept of space, such a position is defined by a set of spatial coordinates. This is not satisfactory, because many different coordinate systems can be defined, between which the choice seems arbitrary. One needs to have a theoretical framework that give a precise meaning to the concept of the space associated with a reference fluid/network. Only a concept of ``spatial tensor" had been defined, to our knowledge. Namely, a spatial tensor at \Couleur{$X \in \mathrm{V}$} was defined as a {\it spacetime tensor} which equals its projection onto the hyperplane \Couleur{$\mathrm{H}_X \equiv v(X)^\perp$}  \cite{Massa1974a, JantzenCariniBini1992}. This is not a very straightforward definition. In addition, a number of time derivatives along a trajectory can then be introduced \cite{ JantzenCariniBini1992}. It is difficult to choose among them.\\

In this conference paper, we will first recall the results obtained previously \cite{A16, A44} regarding the definition of a local reference frame and the associated local space. Then we will announce results of a current work, that aims at defining global notions and at relating them to the formerly introduced local notions.

\section{A local definition of a reference frame and the associated space}


Defining a ``reference fluid" through its 4-velocity field is correct but unpractical. On the other hand, fixing a ``reference system" by the data of a chart \cite{L&L,Moeller1952} is practical, but one may ask: what is physical here? Is there an associated space? What if we change the chart?

\subsection{Space associated with a reference fluid: a sketch}

The three-dimensional space manifold \Couleur{$\mathrm{N}$} associated with a reference fluid (network of observers) can be introduced as {\it the set of the world lines of the observers of the network} \cite{A16}. Thus an {\it element} (point) of \Couleur{$\mathrm{N}$} is a {\it line} of the spacetime manifold \Couleur{$\mathrm{V}$}. Spatial tensor fields are then defined simply as tensor fields on the spatial manifold \Couleur{$\mathrm{N}$} \cite{A16}. At the time of that definition \cite{A16}, the network, hence also \Couleur{$\mathrm{N}$}, was thus defined ``physically", and it was not proved that \Couleur{$\mathrm{N}$} is indeed a differentiable manifold. Nevertheless, it was noted that the spatial metric defined in Refs. \cite{L&L} and \cite{Moeller1952} endows this manifold \Couleur{$\mathrm{N}$} with a time-dependent Riemannian metric, thus with a one-parameter family of metrics. Then, just one time derivative along a trajectory appears naturally \cite{A16}, precisely because we have now just a time-dependent spatial metric tensor instead of a general spacetime metric. This allowed us to unambiguously define {\it Newton's second law} in a general spacetime.


\subsection{A local definition of a reference frame}\label{F definition}

One may define a reference frame as being an {\it equivalence class of charts} which are all defined on a given open subset \Couleur{$\mathrm{U}$} of the spacetime \Couleur{$\mathrm{V}$} and are related two-by-two by a {\it purely spatial} coordinate change: 
\be\label{purely-spatial-change}
\Couleur{x'^0=x^0,\quad x'^k=\phi ^k((x^j))}.
\ee
This does define an equivalence relation \cite{A44}. Thus a reference frame \Couleur{$\mathrm{F}$}, i.e. an equivalence class for this relation, can indeed be given by {\it the data of one chart} \Couleur{$\chi: X \mapsto (x^\mu)$} {\it with its domain of definition} \Couleur{$\mathrm{U}$} (an open subset of the spacetime manifold \Couleur{$\mathrm{V}$}). Namely, \Couleur{$\mathrm{F}$} is the equivalence class of \Couleur{$(\chi ,\mathrm{U})$}. I.e., \Couleur{$\mathrm{F}$} is the set of the charts \Couleur{$\chi'$} which are defined on \Couleur{$\mathrm{U}$}, and which are such that the transition map \Couleur{$f\equiv \chi '\circ \chi ^{-1}\equiv (\phi^\mu )$} corresponds with a purely spatial coordinate change (\ref{purely-spatial-change}).


\subsection{The associated space}

The former definition has physical meaning: the data of a reference frame \Couleur{$\mathrm{F}$} determines the world lines (each of which is included in the common chart domain \Couleur{$\mathrm{U}$}):
\be\label{world line}
 \Couleur{x^j=\mathrm{Constant}\ (j=1,2,3),\ x^0\ \mathrm{variable}}.
\ee
The set of these world lines, as \Couleur{${\bf x}\equiv (x^j)$} varies, is indeed a three-dimensional network. If the charts obey the admissibility condition \Couleur{$g_{00}>0$}, these are time-like world lines. The corresponding 4-velocity field \Couleur{$v$} or rather \Couleur{$v_\mathrm{F}$} is then given by (\ref{Vmu}). The world lines (\ref{world line}) as well as the field \Couleur{$v_\mathrm{F}$} are invariant under the ``internal changes" (\ref{internal-change}). Hence, they are a fortiori invariant under the purely spatial coordinate changes (\ref{purely-spatial-change}). The space \Couleur{$\mathrm{M}=\mathrm{M}_\mathrm{F}$} (in a further step to be equipped with a structure of differentiable manifold) is mathematically defined as the set of the world lines (\ref{world line}).\\

In full detail: let \Couleur{$P_S: \mathbb{R}^4 \rightarrow \mathbb{R}^3,\quad {\bf X}\equiv (x^\mu )\mapsto {\bf x}\equiv (x^j )$}, be the spatial projection. A world line \Couleur{$l$} is an element of the set \Couleur{$\mathrm{M}_\mathrm{F}$} iff there is a chart \Couleur{$\chi \in \mathrm{F}$} and a triplet \Couleur{${\bf x} \equiv (x^j ) \in \mathbb{R}^3$}, such that \Couleur{$l$} is the set, assumed non-empty, of {\it all} points \Couleur{$X$} in the domain \Couleur{$\mathrm{U}$}, whose spatial coordinates are \Couleur{${\bf x}$}:\\
\be\label{l-in-M-by-P_S}
\Couleur{ l =\{\,X\in \mathrm{U};\ P_S(\chi (X))={\bf x}\,\}}\quad \mathrm{and}\ \Couleur{l \ne \emptyset}.
\ee 

Note that the lines (\ref{l-in-M-by-P_S}), hence also their set \Couleur{$\mathrm{M}_\mathrm{F}$}, remain invariant, not only under the purely spatial coordinate changes (\ref{purely-spatial-change}), but under any change (\ref{internal-change}). The coordinate changes (\ref{internal-change}) leave the 4-velocity field \Couleur{$v_\mathrm{F}$} invariant, but in general they change the reference frame, say from \Couleur{$\mathrm{F}$} to \Couleur{$\mathrm{F'}$}, since they generally change the time coordinate. In such a case, we have thus \Couleur{$\mathrm{M}_\mathrm{F}=\mathrm{M}_\mathrm{F'}$}.

\subsection{\Couleur{$\mathrm{M}_\mathrm{F}$} is a differentiable manifold: sketch of the proof}\label{Sketch M_F}

Consider a chart \Couleur{$\chi \in \mathrm{F}$}. With any world line \Couleur{$l\in \mathrm{M}_\mathrm{F}$}, let us associate the triplet \Couleur{${\bf x}\equiv (x^j)$} made with the {\it constant} spatial coordinates of the points \Couleur{$X\in l$}. We thus define a mapping
\be\label{def-chi-tilde}
\Couleur{\widetilde{\chi }: \mathrm{M}_\mathrm{F}\rightarrow \mathbb{R}^3,\quad l\mapsto {\bf x}} \mathrm{\ such\ that\ }\Couleur{\forall X \in l,\  \chi^j(X)=x^j\ (j=1,2,3)}.
\ee
Through Eq. (\ref{l-in-M-by-P_S}), the world line \Couleur{$l \in \mathrm{M}_\mathrm{F}$} is determined uniquely by the data \Couleur{${\bf x}$}. I.e., the mapping \Couleur{$\widetilde{\chi }$} is one-to-one. Consider the set \Couleur{$\mathcal{T}$} of the subsets \Couleur{$\Omega \subset \mathrm{M}_\mathrm{F}$} such that
\be\label{def-Topo}
\Couleur{\forall \chi \in \mathrm{F},\quad \widetilde{\chi }(\Omega)} \mathrm{\ is\ an\ open\ set\ in\ }\Couleur{\mathbb{R}^3}.
\ee
One shows that \Couleur{$\mathcal{T}$} is a topology on \Couleur{$\mathrm{M}_\mathrm{F}$}. Then one shows that the set of the mappings \Couleur{$\widetilde{\chi }$} defines a structure of differentiable manifold on that topological space \Couleur{$\mathrm{M}_\mathrm{F}$}: {\it The spatial part of any chart \Couleur{$\chi \in \mathrm{F}$} defines a chart \Couleur{$\widetilde{\chi }$} on \Couleur{$\mathrm{M}_\mathrm{F}$}} \cite{A44}. In particular, the compatibility of any two charts \Couleur{$\widetilde{\chi}$} and \Couleur{$\widetilde{\chi}'$} on \Couleur{$\mathrm{M}_\mathrm{F}$} stems from the fact that any two charts \Couleur{$\chi,\,\chi '$} that belong to one reference frame \Couleur{$\mathrm{F}$} have a common domain \Couleur{$\mathrm{U}$}: since any world line \Couleur{$l \in \mathrm{M}_\mathrm{F}$} is included in \Couleur{$\mathrm{U}$}, one shows easily that \Couleur{$\widetilde{\chi} '\circ \widetilde{\chi}^{-1}=(\phi ^k)$}, the spatial part of the transition map \Couleur{$\chi '\circ \chi ^{-1}$}.

\subsection{Applications of this result}
A Hamiltonian operator of relativistic QM depends {\it precisely} \cite{A42} on the reference frame  \Couleur{$\mathrm{F}$} as defined in Subsect. \ref{F definition}. The Hilbert space \Couleur{$\mathcal{H}$} of quantum-mechanical states is the set of the square-integrable functions defined on the associated {\it space manifold} \Couleur{$\mathrm{M}_\mathrm{F}$} \cite{A43}. Prior to this definition, \Couleur{$\mathcal{H}$} depended on the particular spatial coordinate system. This does not seem acceptable.\\

 The full algebra of spatial tensors can then be defined in a simple way: a spatial tensor field is simply a tensor field on the space manifold \Couleur{$\mathrm{M}_\mathrm{F}$} associated with a reference frame \Couleur{$\mathrm{F}$}. A simple example is the {\it 3-velocity} of a particle (or a volume element) in a reference frame: this is a spatial vector, i.e., the current 3-velocity at an event \Couleur{$X \in \mathrm{U}$} is an element of the tangent space at \Couleur{$l(X)\in \mathrm{M}_\mathrm{F}$}. \{\Couleur{$l(X)$} is the unique line \Couleur{$l\in \mathrm{M}_\mathrm{F}$}, such that \Couleur{$X\in l$} \cite{A44}.\} As another example, the {\it rotation rate of a spatial triad} is an antisymmetric spatial tensor field of the $(0\ 2)$ type \cite{A47}.

\subsection{Questions left open by that result}

These definitions of a {\it reference frame} and the associated space manifold apply to a domain \Couleur{$\mathrm{U}$} of the spacetime \Couleur{$\mathrm{V}$}, such that at least one regular chart can be defined over the whole of \Couleur{$\mathrm{U}$}. Thus these are {\it local} definitions, since in general the whole spacetime manifold \Couleur{$\mathrm{V}$} cannot be covered by a single chart. Whence the questions:\\

 Can the definition of a {\it reference fluid} by the data of a {\it global} four-velocity field \Couleur{$v$} lead to a global notion of space? If yes, what is the link with the former local notions?

\section{The global space manifold \Couleur{$\mathrm{N}_v$} associated with a non-vanishing vector field \Couleur{$v$}}

\vspace{2mm}

Given a global vector field \Couleur{$v$} on the spacetime \Couleur{$\mathrm{V}$}, and given an event \Couleur{$X\in \mathrm{V}$}, let \Couleur{$C_X$} be the solution of 
\be\label{Def C}
\Couleur{\frac{dC}{ds}=v(C(s)),\qquad C(0)=X}
\ee
that is defined on the {\it largest possible} open interval \Couleur{$\mathrm{I}_X$} containing \Couleur{$0$} \cite{DieudonneTome4}. Call the {\it range} \Couleur{$l_X\equiv C_X(\mathrm{I}_X)\subset \mathrm{V}$} the ``maximal integral curve at \Couleur{$X$}". If \Couleur{$X'\in l_X\,$}, then it is easy to show that \Couleur{$l_{X'}=l_X$}.\\

We define the {\it global space \Couleur{$\mathrm{N}_v$} associated with the vector field \Couleur{$\,v\,$}} as the set of the maximal integral curves of \Couleur{$\,v\,$}:
\be
\Couleur{\mathrm{N}_v\equiv \{l_X;\ X\in \mathrm{V}\}}.
\ee

\subsection{Local existence of adapted charts}\label{adapted charts}

A chart \Couleur{$\chi$} with domain \Couleur{$\mathrm{U}\subset \mathrm{V}$} is said {\it \Couleur{$v$}--adapted} iff the spatial coordinates remain constant on any integral curve \Couleur{$l$} of \Couleur{$v$} --- more precisely, remain constant on \Couleur{$l \cap \mathrm{U}$}:
\be\nonumber
\Couleur{\forall l\in \mathrm{N}_v,\ \exists {\bf x}\equiv (x^j)\in \mathbb{R}^3}:
\ee
\be\label{Def adapted}
\qquad \qquad \qquad \Couleur{\forall X\in l\cap \mathrm{U},\quad P_S(\chi (X))={\bf x}}.
\ee

\vspace{1mm}
 For any \Couleur{$v$}--adapted chart \Couleur{$\chi$}, the mapping 
\be\label{Def chi-bar}
\Couleur{\bar{\chi}: l\mapsto {\bf x}} \mathrm{\ \,such\ that\ (\ref{Def adapted})\ is\ verified}
\ee
is well defined on 
\be\label{Def D_U}
\Couleur{\,\mathrm{D}_\mathrm{U}\equiv \{l\in \mathrm{N}_v;\ l\cap \mathrm{U} \ne \emptyset \}}.
\ee 
Call the \Couleur{$v$}--adapted chart \Couleur{$\chi$} {\it nice} if the mapping \Couleur{$\bar{\chi}$} is one-to-one. On the other hand, call a non-vanishing
\footnote{\
Note that a time-like vector field is non-vanishing. However, we don't need that \Couleur{$v$} be time-like.
}
global vector field \Couleur{$v$} {\it normal} if its flow has the following property that, we can indicate convincingly, is true unless \Couleur{$v$} is ``pathological": {\it Any point \Couleur{$X \in \mathrm{V}$} has an open neighborhood \Couleur{$\mathrm{U}$} such that: (i) for any maximal integral curve \Couleur{$l$} of \Couleur{$v$}, the intersection \Couleur{$l\cap \mathrm{U}$} is a connected set, and (ii) there is a chart \Couleur{$\chi $} with domain \Couleur{$\mathrm{U}$}, such that the corresponding natural basis \Couleur{$(\partial _\mu )$} verifies \Couleur{$v=\partial _0$} in \Couleur{$\mathrm{U}$}.} It is easy to prove the following:

\paragraph{Theorem 1.}\label{Theorem1} {\it Let the global non-vanishing vector field \Couleur{$\,v\,$} on \Couleur{$\mathrm{V}$} be normal. Then, for any point \Couleur{$X \in \mathrm{V}$}, there exists a nice \Couleur{$\,v$}--adapted chart \Couleur{$\chi$} whose domain is an open neighborhood of \Couleur{$X$}.}

\subsection{Manifold structure of the global set \Couleur{$\mathrm{N}_v$}}


Consider the set \Couleur{$\mathcal{F}_v$} made of all nice \Couleur{$\,v$}--adapted charts on the spacetime manifold \Couleur{$\mathrm{V}$}, and consider the set \Couleur{$\mathcal{A}$} made of the mappings \Couleur{$\bar{\chi}$}, where \Couleur{$\chi \in \mathcal{F}_v$}, Eq. (\ref{Def chi-bar}). A such mapping \Couleur{$\bar{\chi}$} is defined on the set 
\Couleur{$\,\mathrm{D}_\mathrm{U}$} --- a subset of the three-dimensional ``space" \Couleur{$\mathrm{N}_v$}, Eq. (\ref{Def D_U}). (Here \Couleur{$\mathrm{U}$} is the domain of the \Couleur{$\,v$}--adapted chart \Couleur{$\chi \in \mathcal{F}_v$}.) When \hyperref[Theorem1]{Theorem 1} above applies, we can go further:\\

\noindent First, in exactly the same way as that used \cite{A44} to prove that the set \Couleur{$\mathcal{T}$} (\ref{def-Topo}) is a topology on the ``local" space \Couleur{$\mathrm{M}_\mathrm{F}$}, we can show that the set \Couleur{$\mathcal{T}'$} of the subsets \Couleur{$\Omega \subset \mathrm{N}_v$} such that 
\be\label{def-Topo-prime}
\Couleur{\forall \chi \in \mathcal{F}_v,\quad \bar{\chi }(\Omega)} \mathrm{\ is\ an\ open\ set\ in\ }\Couleur{\mathbb{R}^3},
\ee
is a topology on the global space \Couleur{$\mathrm{N}_v$}. [We define \Couleur{$\bar{\chi }(\Omega)\equiv \bar{\chi }(\Omega\cap \mathrm{D}_\mathrm{U})$}.] \\

\noindent Second, we can show that \Couleur{$\mathcal{A}$} is an atlas on that topological space, thus defining a structure of differentiable manifold on the global set \Couleur{$\mathrm{N}_v$}. In order to show this, the main thing to prove is the compatibility of any two charts \Couleur{$\bar{\chi}$}, \Couleur{$\bar{\chi}'$} on \Couleur{$\mathrm{N}_v$}, associated with two nice \Couleur{$v$}-adapted charts \Couleur{$\chi,\chi' \in \mathcal{F}_v$}.\\

In the case of the space manifold \Couleur{$\mathrm{M}_\mathrm{F}$} associated with a local reference frame \Couleur{$\mathrm{F}$}, the compatibility of two associated charts \Couleur{$\widetilde{\chi}$} and \Couleur{$\widetilde{\chi}'$} on \Couleur{$\mathrm{M}_\mathrm{F}$} was rather easy to prove, see the end of Sect. \ref{Sketch M_F}. In contrast, two \Couleur{$\,v\,$}-adapted charts \Couleur{$\chi $} and \Couleur{$\chi '$} have in general different domains \Couleur{$\mathrm{U}$} and \Couleur{$\mathrm{U}'$} and we may have   
\be
 \Couleur{\mathrm{U} \cap \mathrm{U}'= \emptyset ,\quad l \cap \mathrm{U} \ne \emptyset,\quad l\cap \mathrm{U}'\ne \emptyset}.
\ee
I.e., the domains of the charts \Couleur{$\chi $}  and \Couleur{$\chi ' $} do not overlap, but the domains of the mappings \Couleur{$\bar{\chi}$} and \Couleur{$ \bar{\chi}'$} do. The solution of this difficulty can be sketched as follows. Consider \Couleur{${\bf x} \in \mathrm{Dom}(\bar{\chi}'\circ \bar{\chi}^{-1} )=\bar{\chi}(\mathrm{D}_\mathrm{U}\cap \mathrm{D}_{\mathrm{U}'})$}. Since \Couleur{${\bf x} \in \bar{\chi}(\mathrm{D}_\mathrm{U})$}, \Couleur{$\exists l\in \mathrm{N}_v$} and \Couleur{$\exists X\in l\cap \mathrm{U}$}: \Couleur{${\bf x} = \bar{\chi}(l)=P_S(\chi (X))$}. Let \Couleur{$\chi (X)=(t,{\bf x})$}. We use the flow of the vector field \Couleur{$v$} to associate smoothly with any point \Couleur{$Y$} in some neighborhood \Couleur{$\mathrm{W}\subset \mathrm{U}$} of \Couleur{$X$},  a point \Couleur{$g(Y)\in \mathrm{U}' $}. Then we may write for \Couleur{${\bf y}$} in a neighborhood of  \Couleur{${\bf x} $}:
\be
\Couleur{ \left( \bar{\chi}' \circ \bar{\chi}^{-1} \right) ({\bf y})=P_S(\chi '(g(\chi ^{-1}(t,{\bf y}))))},
\ee
showing the smoothness of \Couleur{$\bar{\chi}' \circ \bar{\chi}^{-1}$}. Using this, we show that the set \Couleur{$\mathcal{A}$} of the mappings \Couleur{$\bar{\chi }$} is an atlas on \Couleur{$\mathrm{N}_v$}, making it a differentiable manifold.

\subsection{The local manifold \Couleur{$\mathrm{M}_\mathrm{F}$} is a submanifold of \Couleur{$\mathrm{N}_v$}}

Let \Couleur{$v$} be a normal non-vanishing vector field on  \Couleur{$\mathrm{V}$}, and let \Couleur{$\mathrm{F}$} be a reference frame {\it made of nice \Couleur{$\,v$}--adapted charts,} all defined on the same open set \Couleur{$\mathrm{U}\subset \mathrm{V}$}. \\

\noi Let \Couleur{$l\in \mathrm{M}_\mathrm{F}$}, thus there is some chart \Couleur{ $\chi\in \mathrm{F}$} and some \Couleur{${\bf x}\in 
\mathbb{R}^3$} such that \Couleur{ $l =\{\,X\in \mathrm{U};\ P_S(\chi (X))={\bf x}\,\}$}. Then, for any \Couleur{$\,X \in l$}, the curve \Couleur{$l_X$} is the same maximal integral curve \Couleur{$l' \in \mathrm{N}_v\,$}, and we have \Couleur{ $\,l=l' \cap\mathrm{U}$}. We have moreover \Couleur{$\,l'=\bar{\chi }^{-1}({\bf x})=\bar{\chi }^{-1}(\widetilde{\chi }(l))$}. Hence, the mapping \Couleur{$I: \mathrm{M}_\mathrm{F}\rightarrow \mathrm{N}_v, \ l\mapsto l'\ $} is just \Couleur{$I=\bar{\chi }^{-1}\circ \widetilde{\chi }$}. This one-to-one mapping of \Couleur{$\mathrm{Dom}(\widetilde{\chi })=\mathrm{M}_\mathrm{F}$} onto \Couleur{$\mathrm{Dom}(\bar{\chi })=\mathrm{D}_\mathrm{U}$} is a diffeomorphism, hence it is an immersion of \Couleur{$\mathrm{M}_\mathrm{F}$} into \Couleur{$\mathrm{N}_v$}. Thus \Couleur{$\mathrm{M}_\mathrm{F}$} is made of the intersections with the local domain \Couleur{$\mathrm{U}$} of the world lines belonging to \Couleur{$\mathrm{N}_v$}, and {\it we may identify the local space \Couleur{$\mathrm{M}_\mathrm{F}$} with the submanifold  \Couleur{$I(\mathrm{M}_\mathrm{F})=\mathrm{D}_\mathrm{U}$} of the global space \Couleur{$\mathrm{N}_v$}}. Now the manifold structure of \Couleur{$\mathrm{N}_v$} entails that, for any nice \Couleur{$v$}-adapted chart \Couleur{$\chi \in \mathcal{F}_v$}, the associated mapping  \Couleur{$\bar{\chi }$} with domain  \Couleur{$\mathrm{D}_\mathrm{U}$} is a chart on the topological space \Couleur{$(\mathrm{N}_v,\mathcal{T}')$}. In turn, this fact involves the statement that \Couleur{$\mathrm{D}_\mathrm{U}$} is more specifically an {\it open subset} of \Couleur{$\mathrm{N}_v$}.

\section{Conclusion}
\vspace{2mm}

A reference frame can be defined as an equivalence class of spacetime charts \Couleur{$\chi $} which have a common domain \Couleur{$\mathrm{U}$} and which exchange two-by-two by a purely spatial coordinate change \cite{A44}. In addition to being mathematically correct, this definition is practical, because it gives a methodology to use coordinate systems in a consistent and physically meaningful way: the data of one spacetime coordinate system \Couleur{$(x^\mu )$} defines (in its domain of definition \Couleur{$\mathrm{U}$}) the 4-velocity field of a network of observers, Eq. (\ref{Vmu}). The coordinate systems that exchange with \Couleur{$(x^\mu )$} by a purely spatial coordinate change (\ref{purely-spatial-change}) belong to the same reference frame and indeed the associated 4-velocity field (\ref{Vmu}) is the same. Using a general coordinate change instead, allows us to go to any other possible reference frame.\\

\noi A precise notion of the physical {\it space} associated with a given reference network did not exist before for a general spacetime, to our knowledge. We defined two distinct concepts: a local one and a global one, which however are intimately related together. In either case, the space is the set of the world lines that belong to the given (local) reference frame, respectively to the given (global) reference fluid: \\

i) Consider a (local) reference frame in the specific sense meant here, i.e. a set \Couleur{$\mathrm{F}$} of charts, all defined on the same subdomain \Couleur{$\mathrm{U}$} of the spacetime, and exchanging by a change of the form (\ref{purely-spatial-change}). This allows one to define a ``local space" \Couleur{$\mathrm{M}_\mathrm{F}$}: this is the set of the world lines (\ref{world line}) [more precisely the set of the world lines (\ref{l-in-M-by-P_S})] \cite{A44}. Each of these world lines is included in the common domain \Couleur{$\mathrm{U}$} of all charts \Couleur{$\chi \in \mathrm{F}$}. \\

ii) The data of a (global) reference fluid, i.e. a global non-vanishing 4-vector field \Couleur{$\,v\,$}, allows one to define a ``global space" \Couleur{$\mathrm{N}_v$}: this is the set of the maximal integral curves of \Couleur{$\,v\,$}. \\

\noi Both of the local space \Couleur{$\mathrm{M}_\mathrm{F}$} and the global space \Couleur{$\mathrm{N}_v$} can be endowed with a structure of differentiable manifold (when \hyperref[Theorem1]{Theorem 1} applies, for the global space). The manifold structure gives a firm status to the space attached to a reference network and allows us to define spatial tensors naturally, as tensor fields on the space manifold. It has also a practical aspect: Locally, the position of a point in the space can be specified by different sets of spatial coordinates, which exchange smoothly: \Couleur{$x'^k=\phi ^k((x^j))\quad (j,k=1,2,3)$}, and we may use standard differential calculus for mappings defined on that space, by choosing any such coordinates. This applies to both the local space \Couleur{$\mathrm{M}_\mathrm{F}$} and the global space \Couleur{$\mathrm{N}_v$}. \\

\noi There is a close link between the local space \Couleur{$\mathrm{M}_\mathrm{F}$} and the global space \Couleur{$\mathrm{N}_v$}, provided the three-dimensional network of observers is indeed the same in the two cases --- i.e., provided that, in any of the charts which make the reference frame \Couleur{$\mathrm{F}$}: (i) any of the integral curves \Couleur{$l\in \mathrm{N}_v$} has constant space coordinates \Couleur{$x^j$}, and (ii) the mapping \Couleur{$l\mapsto (x^j)$} is one-to-one.
If that is true, one may associate with each world line \Couleur{$l\in \mathrm{M}_\mathrm{F}$} the world line \Couleur{$l'\in \mathrm{N}_v$}, of which \Couleur{$l$} is just the intersection with the domain \Couleur{$\mathrm{U}$}. Thus the local space can be identified with an open subset of the global space.

\end{document}